\documentclass{article}
\usepackage{geometry,cite}  
\usepackage{amsmath,amssymb}

\begin{document}

\begin{center}

{\Large\bf Dipolar induced anisotropy in the random-field Ising model}

\bigskip{} {\large L. L. Afremov$^a$ and A. V. Panov$^{b,}$\footnote{{\it\
E-mail:} panov@iacp.dvo.ru}

}

\bigskip{}

{\it
$^a$Physics Department, Far-Eastern State University, 8, Sukhanova st.,
Vladivostok, 690600, Russia

$^b$Institute for Automation and Control Processes, 
Far-Eastern Branch of Russian Academy of Sciences, \\
5, Radio st., Vladivostok, 690041, Russia}

\end{center}

\begin{abstract} 

The crystalline ferromagnet with dipole-dipole interactions is studied within
the framework of the three-dimensional random-field Ising model. The field
distribution function and the analytic expression for temperature dependent
free energy are found. The dipolar induced magnetic anisotropy energy of bulk
cubic ferromagnetic materials is calculated.

\end{abstract}

In recent years several papers describing the magnetic anisotropy induced by
the dipole-dipole interactions in ultrathin films were published
\cite{hucht97,hucht99,levy}. In Refs.~\cite{hucht97,hucht99} the system
consisting of several ferromagnetic monolayers was considered in the framework
of a classic Heisenberg model. The temperature driven transition from
out-of-plane to in-plane magnetization was explained. In Ref.~\cite{levy} the
magnetic anisotropy of the two-dimensional lattice was analyzed by means of the
dipole-dipole interactions.

The random-field method in the Ising model \cite{bel92} allows us to add the
dipolar interaction and to estimate induced magnetic anisotropy.

\section{Model}

Consider $N+1$ spins positioned at the sites of crystalline ferromagnet lattice.
Let us suppose that one of the site is located at the origin of the coordinates.
The probability of finding the interaction field $\mathbf{H}$ to lie in the
interval $(\mathbf{H},\mathbf{H}+d\mathbf{H})$ at the origin is \cite{bel92,bel85}
\begin{displaymath}
\delta \left[\mathbf{H} - \sum^N_{k = 1}\mathbf{h}_{k} \left( \mathbf{m}_k
,\mathbf{r}_k \right) \right] d \mathbf{H} ,
\end{displaymath} 
where $\mathbf{h}_{k} \left( \mathbf{m}_k,\mathbf{r}_k \right)$ is interaction
field induced by the atom with the magnetic moment $\mathbf{m}_k$ situated at
the site $\mathbf{r}_k$, $\delta$ is Dirac's delta function. Then the random
field distribution function will be
\begin{equation}
W (\mathbf{H} ) d \mathbf{H} = \int \delta \left[\mathbf{H} - \sum^N_{k =
1}\mathbf{h}_k \left( \mathbf{m}_k ,\mathbf{r}_k \right) \right]\prod^N_{k =
1} \tau_k\left( \mathbf{m}_k \right) d \mathbf{m}_{k} d \mathbf{H} ,
\end{equation} 
where $\tau_k \left( \mathbf{m}_k \right)$ is the distribution over atom
magnetic moments $\mathbf{m}_k$ function. In the approach of the Ising model,
when the magnetic moments are aligned with the selected direction given by unite 
vector $\mathbf{s}$ and their
magnitudes are equal to $m_0$, $\tau_k$ takes the following form
\begin{displaymath}
\tau_k \left( \mathbf{m}_k \right) = \alpha_k \delta \left( m_0\mathbf{s}
-\mathbf{m}_k \right) + \beta_k \delta \left( m_0\mathbf{s} +\mathbf{m}
_k \right) ,
\end{displaymath} 
where $\alpha_k$ and $\beta_k$ are the probabilities of finding magnetic moment
to be oriented along and opposite to chosen direction. Let us consider a case
where the interaction field $\mathbf{h}_{k} \left( \mathbf{m}_k,\mathbf{r}_k \right)$
is a sum of exchange and dipole-dipole interaction:
\begin{equation}
\mathbf{h}_k \left( \mathbf{m}_k ,\mathbf{r}_k \right)= J ( r_k ) 
\frac{\mathbf{m}_{k}}{m_k}
 + \frac{3 (\mathbf{r}_k
\cdot \mathbf{m}_k )\mathbf{r}_k - r^2_k\mathbf{m}_k}{r^5_k} 
\label{hk}, 
\end{equation} 
where $J ( r_k )=J$ being exchange constant if $k$th atom is nearest neighbor 
and $J ( r_k )=0$
otherwise.

Let us replace  $\alpha_k$ 
and $\beta_k$ by their average values $\bar\alpha$ and $\bar\beta$ which are
given by the equation \cite{bel92}
\begin{equation}
\bar{\alpha} = \int \frac{\exp \left( m_0\mathbf{s} \cdot \mathbf{H} / k_B T
\right)}{2 \cosh \left( m_0\mathbf{s} \cdot \mathbf{H} / k_B T \right)} W
(\mathbf{H} ) d \mathbf{H} ,\quad  \bar{\alpha} + \bar{\beta} = 1,
\label{alpha}\end{equation} 
where $k_B$ is Boltzmann constant, $T$ is
temperature.

The characteristic function
$A (\boldsymbol{\rho} ) = \int W (\mathbf{H} ) \exp ( i \boldsymbol{\rho} \cdot
\mathbf{H} ) d \mathbf{H}$, being the Fourier transform of the function
$W (\mathbf{H} )$, takes the form
\begin{equation}
A (\boldsymbol{\rho} ) = \prod_k \left[\bar{\alpha} \exp \left( i\boldsymbol{\rho}
\cdot \mathbf{h}_k ( m_0\mathbf{s} ,\mathbf{r}_k ) \right) + \bar{\beta}
\exp \left( - i\boldsymbol{\rho} \cdot \mathbf{h}_k ( m_0\mathbf{s} ,\mathbf{r}
_k ) \right) \right] .
\end{equation} 
After the inverse transformation the distribution function is
\begin{equation}
W (\mathbf{H} ) = \sum_{j_1 = 0}^1 C_1^{j_1} \sum_{j_2 = 0}^1 C_1^{j_2}
\cdots \sum_{j_N = 0}^1 C_1^{j_N} \bar{\alpha}^{\sum_{k = 1}^N j_k}
\bar{\beta}^{N - \sum_{k = 1}^N j_k} \delta \left[\mathbf{H} - \sum_{k =
1}^N ( 2 j_k - 1)\mathbf{h}_k ( m_0\mathbf{s} ,\mathbf{r}_k ) \right] ,
\end{equation} 
where $C_1^{j_k}=1$ are binomial coefficients.

One can obtain the free energy per atom using relation
$F = - k_B T \ln \int \exp \left( \frac{m_0\mathbf{s} \cdot \mathbf{H}}{k_B T}
\right) W (\mathbf{H} ) d \mathbf{H}$. After integrating $F$ will be
\begin{equation}
F = - k_B T \sum_k \ln \left[\bar{\alpha}
 \exp \left( \frac{m_0\mathbf{s} \cdot \mathbf{h}_k ( m_0\mathbf{s}
,\mathbf{r}_k )}{k_B T} \right) + \bar{\beta} \exp \left( -
\frac{m_0\mathbf{s} \cdot \mathbf{h}_k ( m_0\mathbf{s} ,\mathbf{r}_k )}{k_B
T} \right) \right] \label{free}
\end{equation} 
and substituting (\ref{hk}) into Eq.~(\ref{free}) $F$ takes the form
\begin{equation}
F = - k_B T \sum_k \ln \left\{\bar{\alpha} \exp \left[\frac{1}{k_B T}
\left( J ( r_k ) + m_0^2\frac{3 (\mathbf{r}_k \cdot \mathbf{s} )^2 - r^2_{^{}
k}\mathbf{}}{r^5_k} \right) \right] + \bar{\beta} \exp \left[-
\frac{1}{k_B T} \left( J ( r_k ) + m_0^2\frac{3 (\mathbf{r}_k \cdot
\mathbf{s} )^2 - r^2_{^{} k}\mathbf{}}{r^5_k} \right) \right] \right\} .
\label{freesubs}
\end{equation} 
Note that the terms describing contribution of the external field 
$\mathbf{H}_\mathrm{ext}$ and single-ion anisotropy
$- m_0\mathbf{s} \cdot \mathbf{H}_\mathrm{ext} + D ( s_x^4 + s_y^4 + s_z^4 )$,
$D$ being single-ion cubic anisotropy constant, may be added to the 
Eq.~(\ref{freesubs}). One can calculate $\bar\alpha$ and $\bar\beta$
using equation derived from (\ref{alpha}) \cite{bel92} taking account of the
exchange coupling only:
\begin{equation}
\bar{m} = 2 \bar{\alpha} - 1 = \sum_{j = 0}^Z C_Z^j \bar{\alpha}^j ( 1 -
\bar{\alpha} )^{Z - j} \tanh \left[\frac{J}{k_B T} ( 2 j - Z) \right] , 
\end{equation} 
where $\bar{m} = \bar{\alpha} - \bar\beta$, $Z$ is number of nearest neighbors.

\section{Results}

Let us treat dipole interaction as small parameter with respect to exchange
coupling ($m_0^2\ll J$), also consider the case of high temperatures ($m_0^2\ll
k_B T$). Expand $F$ into a series in powers of $m_0^2$ and direction cosines
$s_x, s_y, s_z$. Taking the identity for cubic crystals 
\begin{displaymath}
s_x^4 + s_y^4 + s_z^4 = 1 - 2 \left( s_x^2 s^2_y + s_y^2 s^2_z +
s_z^2 s^2_x \right) 
\end{displaymath}
into account and retaining first nonzero anisotropic terms of this expansion
we will obtain the anisotropic part of free energy
\begin{equation}
F_{\mathrm{an}} = - 36 \bar{\alpha} \bar{\beta} \frac{m_0^4}{k_B T}
\frac{1}{( a / 2 )^6} \left[3 L_1 - L_2 + L_3 \frac{1}{\bar{\alpha} \exp
\left( J / k_B T \right) + \bar{\beta} \exp \left( - J / k_B T
\right)} \right] \left( s_x^2 s^2_y + s_y^2 s^2_z + s_z^2
s^2_x \right) ,
\end{equation} 
$a$ being lattice constant,
\begin{equation}
L_1 = \sum^{\infty}_{\begin{array}{c} \scriptstyle j , k , l = - \infty \\ \scriptstyle 
j^2 + k^2 + l^2 \geqslant 4\end{array}} \frac{j^4}{( j^2 + k^2 + l^2 )^5} ,
\end{equation} 
\begin{equation}
L_2 = \sum^{\infty}_{\begin{array}{c} \scriptstyle j , k , l = - \infty \\ \scriptstyle 
j^2 + k^2 + l^2 \geqslant 4\end{array}} \frac{j^2 k^2}{( j^2 + k^2 + l^2 )^5} ,
\end{equation} 
can be calculated numerically.
For bcc lattice it was calculated that $L_1\approx 0.00454$, $L_2\approx 
0.04343$, $L_3=\frac{16}{243}$ and for fcc lattice $L_1\approx 0.03200$, $L_2\approx 
0.12912$, $L_3=\frac{1}{8}$.

As one can see at $T=0$ ferromagnet is totally ordered and $F_{\mathrm{an}}$
derived from Eq.~(\ref{freesubs}) is equal to zero since quadrupole moment of
this system is zero. With increase in temperature $|F_{\mathrm{an}}|$ raises
and at  $T>T_c$ $F_{\mathrm{an}}\neq 0$ because of used approximation of Ising
model (magnetic moments of all atoms are parallel to chosen direction).

The method proposed can be used for estimation of the magnetic anisotropy
induced by dipole-dipole interactions in two-dimensional and bulk materials.

This work was supported by Russian Ministry of Education. The authors would
like to thank V. I. Belokon' for fruitful discussion.

\end{document}